\documentclass[aps,prx,reprint,twocolumn,groupeaddress,superscriptaddress,longbibliography]{revtex4-2}
\usepackage[dvipdfmx]{graphicx}
\usepackage{helvet}
\usepackage{amsmath,amssymb}
\usepackage{bm}
\usepackage{colortbl}
\usepackage{braket}
\begin{document}


\title{Chirality-Induced Electrical Generation of Magnetism in Nonmagnetic Elemental Tellurium}

\author{Tetsuya Furukawa}
\email{tf@imr.tohoku.ac.jp}
\affiliation{Institute for Materials Research, Tohoku University, Sendai 980-8577, Japan}
\affiliation{Department of Applied Physics, Tokyo University of Science, Tokyo 125-8585, Japan}

\author{Yuta Watanabe}
\affiliation{Department of Applied Physics, Tokyo University of Science, Tokyo 125-8585, Japan}

\author{Naoki Ogasawara}
\affiliation{Department of Applied Physics, Tokyo University of Science, Tokyo 125-8585, Japan}

\author{Kaya Kobayashi}
\affiliation{Research Institute for Interdisciplinary Science, Okayama University, Okayama 700-8530, Japan}

\author{Tetsuaki Itou}
\email{tetsuaki.itou@rs.tus.ac.jp}
\affiliation{Department of Applied Physics, Tokyo University of Science, Tokyo 125-8585, Japan}

\date{\today}

\begin{abstract}
Chiral matter has a structure that lacks inversion, mirror, and rotoreflection symmetry; thus, a given chiral material has either a right- or left-handed structure.
In chiral matter, electricity and magnetism can be coupled in an exotic manner beyond the classical electromagnetism (e.g., magneto chiral effect in chiral magnets). 
In this paper, we give a firm experimental proof of the linear electric-current-induced magnetization effect in bulk nonmagnetic chiral matter elemental trigonal tellurium. 
We measured a $^{125} $Te nuclear magnetic resonance (NMR) spectral shift under a pulsed electric current for trigonal tellurium single crystals.
We provide general symmetry considerations to discuss the electrically (electric-field- and electric-current-) induced magnetization and clarify that the NMR shift observed in trigonal tellurium is caused by the linear current-induced magnetization effect, not by a higher-order magnetoelectric effect.
We also show that the current-induced NMR shift is reversed by a chirality reversal of the tellurium crystal structure.
This result is the first direct evidence of crystal-chirality-induced spin polarization, which is an inorganic-bulk-crystal analogue of the chirality-induced spin selectivity in chiral organic molecules.
The present findings also show that nonmagnetic chiral crystals may be applied to spintronics and coil-free devices to generate magnetization beyond the classical electromagnetism.
\end{abstract}


\maketitle
\section{Introduction}
In condensed matters, various crystal symmetries and low-energy excitations allow an exotic coupling between electricity and magnetism beyond the classical electromagnetism. These cross-correlation phenomena, called magnetoelectric effects \cite{Dzyaloshinskii1959} (in a broad sense), may have applications in information and energy technology. Among various magnetoelectric effects, the linear electric-field-induced magnetization effect can occur only in a system that lacks both inversion and time-reversal symmetry \footnote{More precisely, in addition to inversion and time-reversal symmetry, some combinations of other crystal symmetries---(proper and improper) rotation, translation, and nonsymmorphic symmetry---and their anti-unitary counterparts can also forbid the linear electric-field-induced magnetization. Consequently, only 58 of 122 magnetic point groups allow the linear electric-field-induced magnetization effect~\cite{Rivera2009}.}. Thus, this magnetoelectric effect is forbidden in nonmagnetic materials. (The term ``magnetization'' is defined in this paper as net macroscopic magnetization, unless there is particular attention.) In contrast, an electric current can induce magnetization in a noncentrosymmetric material, even if  the material has time-reversal symmetry. This current-induced magnetization effect has so far been studied in surface/interface Rashba systems~\cite{ Ganichev2002,Kato2004,Sih2005, Silov2005,Yang2006,Stern2006,Wilamowski2007,Koehl2009,Zhang2014,Ganichev2014} as the Edelstein effect~\cite{Edelstein1990}. In a microscopic viewpoint, an electric current causes an imbalance between opposite spin populations in spin-split bands caused by the breaking of inversion symmetry and the spin--orbit interaction (SOI). A promising direction of the current-induced magnetization effect is to expand the phenomenon into bulk materials. The availability of various types of crystal symmetry in bulk materials allows us to arbitrarily design a relation between the directions of an electric current and induced magnetization. 

A sign of the bulk current-induced magnetization has been observed as current-induced modulation of optical activity in elemental trigonal tellurium ~\cite{Vorobev1979,Shalygin2012}. We recently reported the possibility of the bulk linear current-induced magnetization in trigonal tellurium by observing a hyperfine shift, which is proportional to the local electronic magnetization, of the nuclear magnetic resonance (NMR) spectrum induced by an electric current~\cite{Furukawa2017}. The induced magnetization can be parallel to an applied electric current in tellurium because of its chiral nature, which contrasts sharply with the current-induced magnetization perpendicular to an applied current in Rashba systems. Thus, the possible current-induced magnetization effect in tellurium may provide coil-free devices to generate magnetization and a magnetic field. Moreover, the possible current-induced spin polarization in chiral tellurium may be a bulk-crystal counterpart of the chirality-induced spin selectivity studied in chiral organic molecules~\cite{Naaman2012}. However, the analysis and the argument of our previous work were not perfect, as will be explained later. An incontrovertible proof of this phenomenon is desired.

We prove herein the linear current-induced magnetization effect in chiral matter tellurium in a more stringent manner than the previous report~\cite{Furukawa2017}, providing symmetry considerations to discuss the general magnetoelectric effect and analyzing the observed electrically (electric-field- and electric-current-) induced NMR spectral shifts from the viewpoint of the symmetry considerations. We address two issues. First, we prove that the electrically induced NMR shift is attributable to the linear current-induced magnetization effect, not to higher-order magnetoelectric effects. Second, we present that the chirality reversal of the crystal structure causes the polarity reversal of the current-induced magnetization using a right- and a left-handed single crystal. These two results give a conclusive understanding of the magnetoelectric effect in trigonal tellurium.

\section{Symmetry considerations}
\begin{figure*}[t]
\includegraphics[]{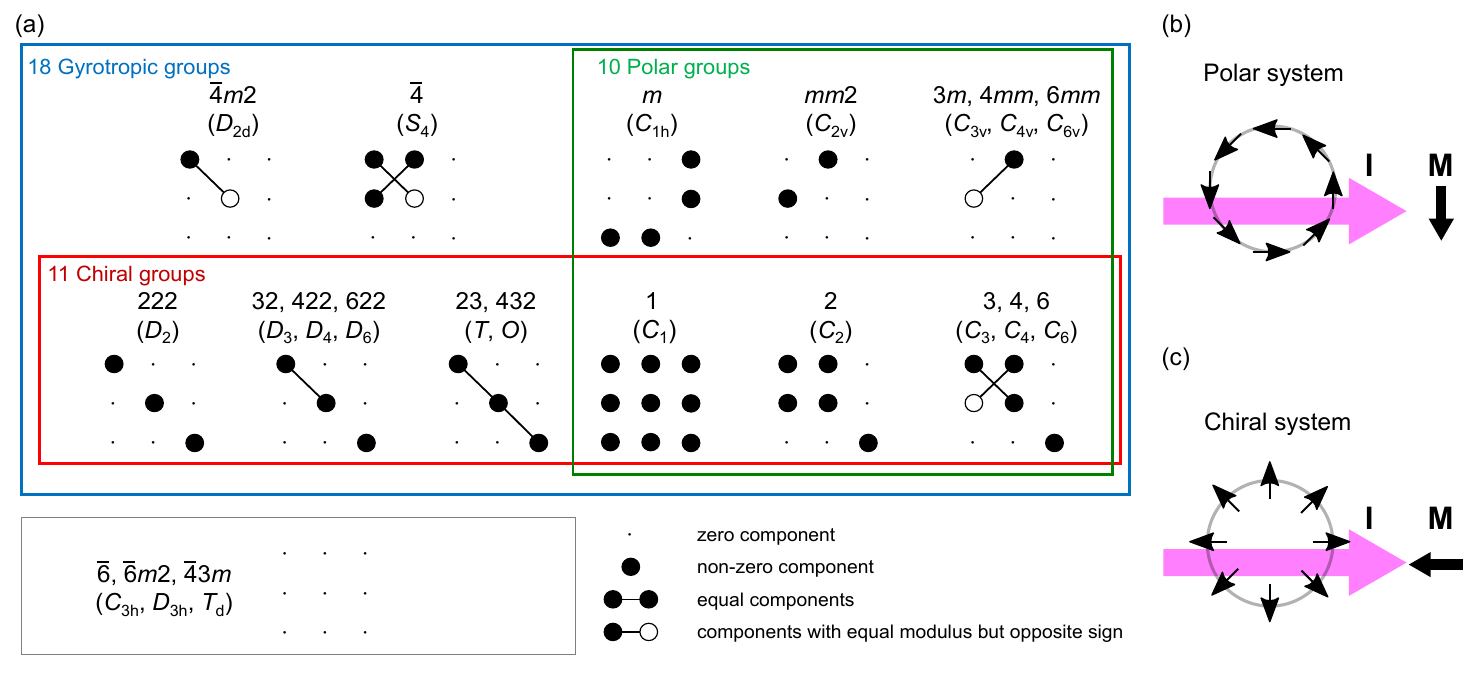}
\caption{Relation between the linear current-induced magnetization and the crystal point groups. (a) Possible second-rank axial tensor forms of the linear current-induced magnetization effect for 21 noncentrosymmetric crystal point groups. Only 18 gyrotropic groups can have nonzero components of the tensor. Three of the noncentrosymmetric groups are non-gyrotropic. The tensor forms for the point groups of the monoclinic systems, $2$($C_{2}$) and $m$($C_{1\mathrm{h} }$), are represented in the first setting, where the two-fold rotation axis (the mirror plane) is parallel (perpendicular) to the $z$ axis. (b) Circular angular-momentum texture in the $\mathbf{k} $ space in a polar system. The current-induced magnetization $\mathbf{M} $ tends to be perpendicular to an applied electric current $\mathbf{I} $. (c) Radial angular-momentum texture in a chiral system. $\mathbf{M} $ tends to be parallel to $\mathbf{I} $.}
\label{Fig. 1}
\end{figure*}
The bulk linear current-induced magnetization is written as $\mathbf{M} = \beta \mathbf{I} $. The second-rank current-induced magnetization tensor $\beta $ must be axial because the tensor connects magnetization $\mathbf{M} $ (an axial vector) and electric current $\mathbf{I} $ (a polar vector). Axial tensors acquire a minus sign through improper rotation; hence, the matrix elements of $\beta $ are more strongly constrained by improper rotation symmetry (e.g., inversion, mirror, and rotatory reflection symmetry) than by proper rotation symmetry, sometimes being zero. For example, inversion symmetry forces all elements of $\beta $ to be zero. Figure~\ref{Fig. 1}a shows the possible tensor forms of $\beta $ for 21 noncentrosymmetric crystal point groups. Notably, all elements of $\beta $ are zero for $\bar{6}$($C_{3\mathrm{h} }$), $\bar{6}m2$($D_{3\mathrm{h} }$), and $\bar{4}3m$($T_{\mathrm{d} }$), even though they do not possess inversion symmetry. Thus, the bulk linear current-induced magnetization effect can occur only in systems belonging to the remaining 18 noncentrosymmetric crystal point groups, called the gyrotropic groups. In the gyrotropic groups, the two sub-classes of polar and chiral groups are important. Polar systems, such as the Rashba systems, have a circular angular-momentum texture in the wavenumber $\mathbf{k} $ space and anti-symmetric (off-diagonal) components in $\beta $; thus, an electric current and an induced magnetization tend to be orthogonal independently of the current direction (Fig.~\ref{Fig. 1}b). Chiral systems (e.g., trigonal tellurium), which have no improper rotation symmetry, have a radial angular-momentum texture and isotropic diagonal components in $\beta$; thus, an electric current and an induced magnetization tend to be parallel (Fig.~\ref{Fig. 1}c).

\section{Crystal and Electronic structures of trigonal Tellurium}
\begin{figure*}[t]
\begin{center}
\includegraphics[]{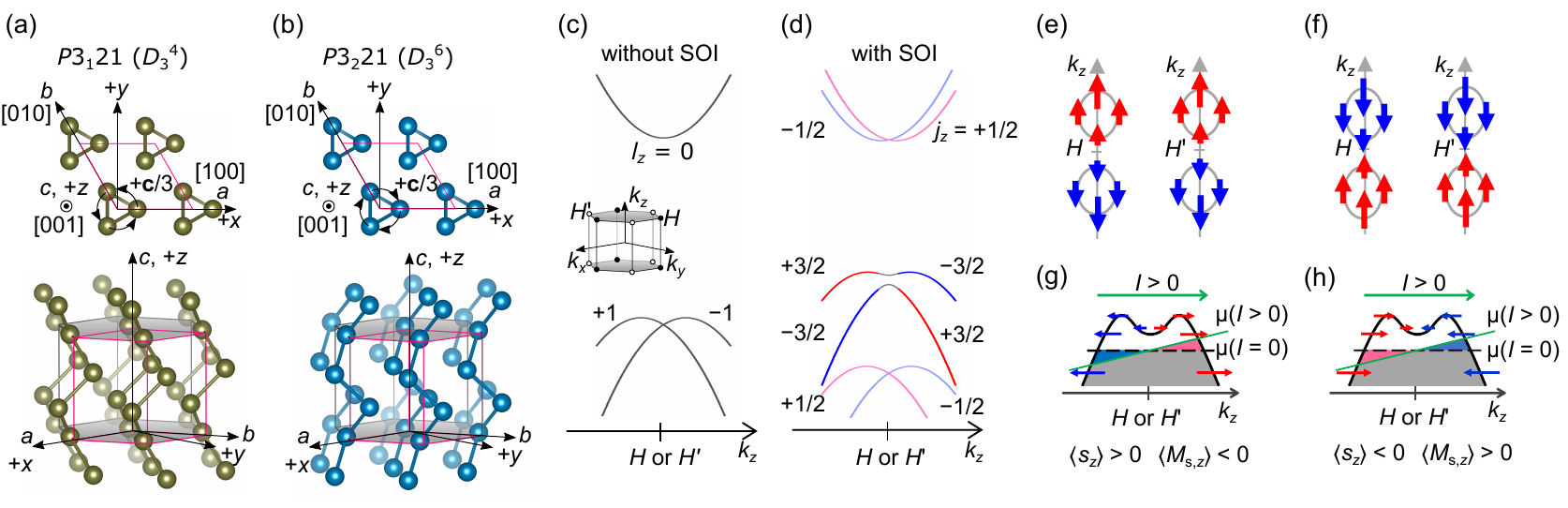}
\caption{
Crystal and electronic structure of trigonal tellurium. (a, b) Crystal structures of tellurium for $P3_{1}21$($D_{3}^{4} $) (a) and $P3_{2}21$($D_{3}^{6}$) (b) with right-handed covalent-bond helices and left-handed helices, respectively. The pink parallelogram indicates a unit cell. A right-handed crystal has the screw operation of $\{\rho _{3}|\mathbf{c} /3\}$ as a symmetry operation, while a left-handed crystal has the operation of $\{\rho _{3}|2\mathbf{c} /3\}$, where $\rho _{3}$ is a $120^{\circ}$ counter-clockwise rotation along the $c$ axis, and $\mathbf{c} $ is the primitive vector along the $c$ direction. (c, d) Schematic band structures along $k_{z}$ (and $k_{x}$ = $k_{y}$ = 0) around the $H$ and $H'$ points with (d) and without (c) the spin--orbit interaction (SOI). The signs of the (inter-atomic) orbital angular momentum ($l_{z}$) and the total angular momentum ($j_{z}$) for each wave number $\mathbf{k} $ and for each band are reversed when the crystal structure chirality is reversed. The bands in the panel (c) are totally spin-degenerate. The hexagonal prism in (c) shows a schematic image of the Brillouin zone of trigonal tellurium. (e, f) Schematic outward (e) and inward (f) angular-momentum texture around the $H$ and $H'$ points when chemical potential crosses the camel-back structure of the highest valence band. (g, h) Relation among the polarities of the net spin-angular momentum $\langle s_{z} \rangle$ and of the net spin magnetization $\langle M_{s,z} \rangle$ and an imbalance between the opposite spin populations for the highest valence band under a positive electric current ($\mathbf{I} \parallel +\mathbf{\hat{k}}_{z}$) for outward (g) and inward (h) spin-angular-momentum textures.
}
\label{Fig. 2}
\end{center}
\end{figure*}
Trigonal tellurium is a semiconductor with a narrow bandgap of $E_{\mathrm{g}}$ = 0.32 eV, with only the $p$-type being available. trigonal tellurium has a crystal structure belonging to either space group $P3_{1}21$($D_{3}^{4} $) or $P3_{2}21$($D_{3}^{6} $) (Figs.~\ref{Fig. 2}a and b). Tellurium atoms form dominant covalent bonds in a helix structure with a three-fold screw symmetry. The helices along the crystal $c$ axis further form a hexagonal arrangement with van der Waals interactions, thereby yielding a chiral crystal structure. Note that a crystal with right-handed helices of the covalent bonds belongs to the space group with a right-handed symmetry element $P3_{1}21$($D_{3}^{4} $), and a crystal with left-handed helices belongs to $P3_{2}21$($D_{3}^{6}$) \footnote{In principle, the chirality of the helices of dominant chemical bonds and that of the space-group screw axes are not always the same (e.g., $\beta $-quartz.)}. In this paper, we define the chirality of the crystal structure by referring to that of the covalent-bond helices. We use the phrases ``right-handed tellurium'' for $P3_{1}21$($D_{3}^{4} $)-tellurium and ``left-handed tellurium'' for $P3_{2}21$($D_{3}^{6} $)-tellurium~\cite{SM}. The chirality of the covalent-bond helices determines the polarities of many physical properties, such as the angular-momentum texture in the $\mathbf{k} $ space ~\cite{Hirayama2015,Tsirkin2018,Sakano2020}, natural optical rotatory power~\cite{Nomura1960,Fukuda1975,Ades1975,Stolze1977}, second-harmonic generation~\cite{Cheng2019}, piezoelectricity~\cite{Arlt1969,Royer1979}, shapes of etch pits~\cite{Koma1970}, resonant diffraction with circularly polarized x rays~\cite{Tanaka2010,Tanaka2012}, polarized neutron scattering~\cite{Brown1996}, and NMR chemical shift~\cite{Koma1973a}  (see Supplemental Material~\cite{SM}).

Trigonal tellurium has the bottoms of conduction bands and the tops of valence bands around the $H$ and $H'$ points~\cite{Betbeder-Matibet1969,Doi1970,Hirayama2015}. These points are at the corners of the hexagonal prism first Brillouin zone and are related to each other by a time-reversal operation; they are not time-reversal invariant momentums. Thus, the time-reversal symmetry of trigonal tellurium does not provide Kramers degeneracy at the $H$ and $H'$ points. Around these points, if spin--orbit interactions are absent, the conduction bands would form a doubly spin-degenerate parabolic band, whereas the valence bands would form spin-degenerate, but (inter-atomic) orbital angular-momentum split bands \footnote{Even in a crystal that lacks full rotational symmetry but has discrete rotational or screw symmetries, the (crystal) orbital angular momentum projected to the rotational or screw symmetry axis can be defined as $l_{z} = m$ mod $n$ for a spin-less system, where integer $m$ appears in the phase of $\pm 2m\pi$ acquired by a wave function through the $n$-time operation of the $n$-fold rotation or the $n$-fold screw operations equivalent to the identity operation, where $+$ and $-$ are for the left- and right-handed symmetry operations, respectively. For $P3_{1}21$($D_{3}^{4} $) right-handed tellurium without the spin--orbit interaction, the degenerate states and the nondegenerate state at the $H$($H'$) point respectively acquire the phases of $\pm 2\pi $ and $0$, through the triple operation of the three-fold \textit{left-handed} screw operation $\{\rho _{3}^{-1} |2\mathbf{c} /3\}$. (Note that the triple operation of the three-fold \textit{right-handed} screw operation $\{\rho _{3}|\mathbf{c} /3\}$, which is another symmetry operation of $P3_{1}21$($D_{3}^{4} $), is not equivalent to the identity operation for the Bloch states at the $H$($H'$) point because it provides the phases of $+3\pi $ and $-\pi $ for the degenerate bands and $+\pi $ for the non-degenerate band.) Hence, the degenerate states and the nondegenerate states at the $H$($H'$) point have $l_{z} = \pm 1$ and $l_{z} = 0$, respectively (Fig.~\ref{Fig. 2}c).} with quadruple degeneracy at the $H$ and $H'$ points owing to three-fold screw and two-fold rotation symmetry (Fig.~\ref{Fig. 2}c).  The spin--orbit interactions lift the spin degeneracy, thereby causing the two spin-split conduction bands crossing at the $H$ and $H'$ points, the highest and second highest nondegenerate valence bands, and the two splitting deeper valence bands with band crossing (Fig.~\ref{Fig. 2}d). This band splitting provides total-angular-momentum textures in the $\mathbf{k} $ space. The two deeper valence bands primally consist of states with total angular momentum $j_{z} = +1/2$ or $-1/2$ and have a hedgehog angular-momentum texture around the $H$ and $H'$ points (i.e., the spin-split conduction bands also have a similar angular-momentum texture~\cite{Hirayama2015}). Notably, the crystal three-fold screw and the twofold rotation symmetries forbid hybridization between the Bloch states with $ j_{z} = +1/2$ and $-1/2$ at the $H$ and $H'$ points, that is, they protect the band crossing. This crossing point is nothing but a Weyl node; thus, the possibility of a topological Weyl semimetal phase is argued for trigonal tellurium ~\cite{Hirayama2015, Nakayama2017, Ideue2019,Sakano2020}. In contrast, the highest and second highest valence bands primally consist of the hybridized states of $ j_{z} = \pm 3/2$, and the wave functions $\ket{j_{\mathrm{z}} = +3/2} \pm \ket{j_{\mathrm{z} } = -3/2}$ are realized at the $H$ and $H'$ points. Such spin-nematic-type hybridization opens a gap between the two valence bands and causes a radial, but almost $k_{z}$-directional spin texture (Figs.~\ref{Fig. 2}e and f). This $c$-axis anisotropy in $p$-type tellurium with a low carrier density causes that only the $c$-axis part of an electric current induces magnetization, which also has only the $c$-axis component. The $\mathbf{k} $ space angular-momentum textures for a right- and a left-handed crystal have opposite polarity, thanks to the axial vector nature of the angular momentum. (Ab initio calculations~\cite{Tsirkin2018,Sakano2020} directly show that the right- [$P3_{1}21$($D_{3}^{4} $)] and left-handed [$P3_{2}21$($D_{3}^{6} $)] crystals have outward and inward angular-momentum textures, respectively (Figs.~\ref{Fig. 2}e and f).) Thus, an electric current should cause a linear current-induced magnetization with polarity depending on the crystal structure chirality (Figs.~\ref{Fig. 2}g and h).

\section{Measuring Method}
We measured the $^{125} $Te NMR spectra for two single crystals (samples \#1 and \#2) at 100 K under an applied pulsed electric current. The crystal structure chirality for each sample was determined by observing the shapes of etch pits on the surfaces of each sample. According to the etch pit study~\cite{Koma1970}, samples \#1 and \#2 are left- [($P3_{2}21$($D_{3}^{6} $)] and right-handed [$P3_{1}21$($D_{3}^{4} $)] tellurium, respectively (Figs.~\ref{Fig. 3}a and ~\ref{Fig. 5}a). We applied a static magnetic field of 8.012 T for Sample \#1 (the left-handed crystal) and 8.327 T for Sample \#2 (the right-handed crystal) almost parallel to the $c$ axis. A pulsed current synchronized with the NMR measurement was also applied almost parallel to the $c$ axis through two electric terminals~\cite{SM}. To avoid movement of the sample by the pulsed electric current, Sample \#1 with a coil for the NMR measurements was coated with epoxy (Stycast 1266) and fixed on a glass plate, and Sample \#2 with a coil was soaked into a pressure medium of Daphne 7373 and placed in a BeCu pressure cell, in which a low pressure of 0.5 GPa was applied at room temperature. The temperature dependence of the two-terminal resistance showed that the present crystals had an impurity hole density less than 1 $\times $ 10$^{16} $ cm$^{-3} $ and that the systems at 100 K were in the extrinsic region with a negligible density of the thermally excited carriers across the band gap. For an arbitrary magnetic field direction, the NMR spectrum of trigonal tellurium should have three lines with the same intensity that correspond to the three atoms in the unit cell. For the present samples, we verified that such a three-line spectrum appeared when the magnetic field direction was rotated away from the $c$ axis, confirming that the samples were single crystals. In the following results, an applied magnetic field was almost parallel to, but slightly tilted from the $c$ axis. The tilting direction was controlled to be precisely aligned to the $y$ axis. Thus, the observed NMR spectra consisted of two lines, a single line and a doubling line, which are respectively associated with one and two of the three tellurium atoms in a unit cell (Figs.~\ref{Fig. 3}a, b and ~\ref{Fig. 5}a). We estimated the angular differences between the magnetic field directions and the $c$ axis to be approximately $7^{\circ}$ for Sample \#1 and $9^{\circ}$  for Sample \#2 by comparing the two-line spectrum with the shift parameters ~\cite{Koma1973a}.

\section{Results}
\subsection{Magnetic-field polarity dependence of current-induced NMR shift}

\begin{figure*}[t]
\includegraphics[]{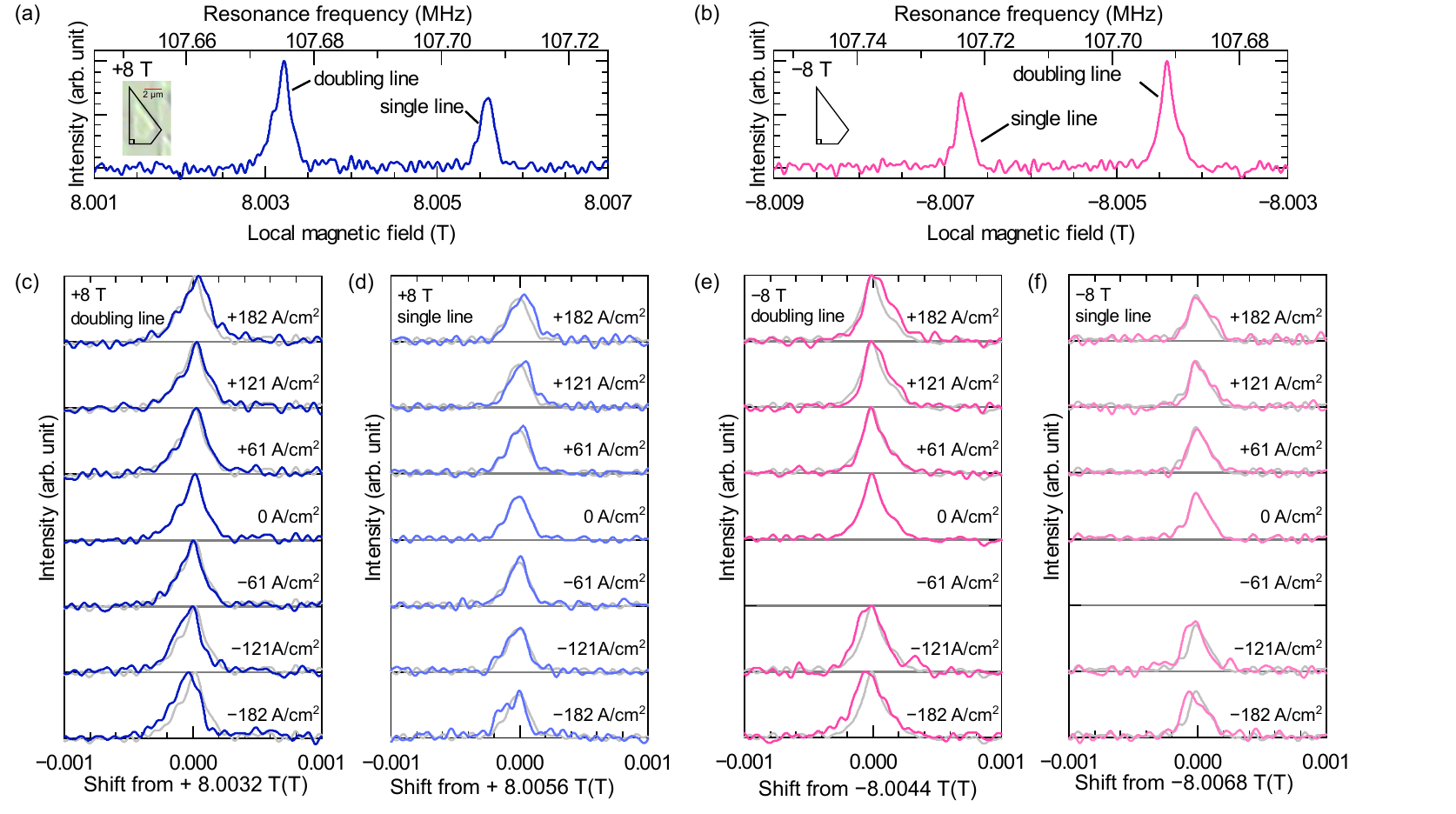}
\caption{
Magnetic-field polarity dependence of the electrically induced NMR shift. 
(a, b) Single-crystal $^{125} $Te-NMR spectra for Sample \#1 for (a) a positive and (b) a negative magnetic field in the absence of a pulsed electric current at 100 K. A magnetic field was applied approximately parallel to the $c$ axis, but slightly tilted to the $y$ axis. The spectra are plotted as a function of the effective magnetic field felt by the $^{125} $Te nuclei~\cite{SM}. The observed NMR spectra consist of two lines, that is, a single line and a doubling line, which are respectively associated with one and two of the three tellurium atoms in a unit cell. The inset in (a) is a photo image of an etch pit on the crystal, whose type appears on the left-handed [$P3_{2}21$($D_{3}^{6} $)] crystal [27]. (c--f) Doubling line (c) and single line (d) in the $^{125} $Te-NMR spectrum for a positive magnetic field and doubling line (e) and single line (f) for a negative magnetic field for different electric current densities. For comparison, the gray lines in each row show the spectrum in the absence of a pulsed electric current. The absence of data for $-$61 Acm$^{-2} $ under a negative field is caused by the irreversible damage of an electric contact of the specimen during the measurement for this condition.
}
\label{Fig. 3} 
\end{figure*}
We observed the electrically induced $^{125} $Te NMR spectral shift in Figs.~\ref{Fig. 3}c--f in more detail than in our previous report~\cite{Furukawa2017}. Below, we show that the electrically induced NMR shift is caused by the linear current-induced magnetization effect, and not by a higher-order magnetoelectric effect. In theoretical viewpoints, the electric-field- and electric-current-induced magnetization can be clearly distinguished. The former is an equilibrium phenomenon without energy dissipation and requires systems that lack both inversion ($P$) and time-reversal ($T$) symmetry. In contrast, the latter is a non-equilibrium phenomenon that requires systems lacking both $P$ and $PT$ symmetry. These relations indicate the duality of the two phenomena: the electric-field-induced magnetization needs an odd-parity component of magnetic ordering in real space, whereas the electric-current-induced magnetization needs an odd-parity component of the magnetization textures (angular-momentum textures) in the $\mathbf{k} $ space. However, in real experiments, the two effects cannot be immediately distinguished because an electric current inevitably needs an electric field. Furthermore, the NMR measurement requires an external magnetic field, which explicitly breaks the time-reversal symmetry of materials and may cause higher-order magnetoelectric effects. Thus, to prove firmly the linear current-induced magnetization effect, it is necessary to rule out the scenarios that the electrically induced magnetization observed originates from the electric-field-induced magnetization and/or higher-order magnetoelectric effects. (This was not done in our previous report~\cite{Furukawa2017}.) From this viewpoint, we first consider the symmetry of the magnetism induced by an electric current $\mathbf{I} $, an electric field $\mathbf{E} $, and a magnetic field $\mathbf{H} $. Phenomenologically, magnetization can be expanded up to the second order of $\mathbf{I} $, $\mathbf{E} $, and $\mathbf{H} $ as follows:
\begin{multline*}
M_{i} = \chi _{ij}H_{j} + \alpha _{ij}E_{j} + \beta _{ij}I_{j} + \Gamma ^{\mathrm{IE}} _{i,jk} I_{j}E_{k} + \Gamma ^{\mathrm{EH}} _{i, jk} E_{j}H_{k} \\
+ \Gamma ^{\mathrm{IH}} _{i,jk} I_{j}H_{k} + \Gamma ^{\mathrm{II}} _{i,jk} I_{j}I_{k} + \Gamma ^{\mathrm{EE}} _{i,jk} E_{j}E_{k} + \Gamma ^{\mathrm{HH}} _{i,jk} H_{j}H_{k},
\end{multline*}
where, $i$, $j$, and $k$ (= $x$, $y$, $z$) are the indices of the matrix elements; $\chi $, $\alpha $, and $\beta $ are the second-rank response tensors for magnetic susceptibility, linear electric-field-induced magnetization, and linear current-induced magnetization, respectively; and $\Gamma ^{\mathrm{IE} } $, $\Gamma ^{\mathrm{EH} } $, $\Gamma ^{\mathrm{IH} } $, $\Gamma ^{\mathrm{II} } $, $\Gamma ^{\mathrm{EE} } $, and $\Gamma ^{\mathrm{HH} } $ are the third-rank response tensors for each bilinear effect. To know which terms can survive under the $P$ and $T$ transformations, we considered the $P$ and $T$ transformations of $\mathbf{M} $, $\mathbf{I} $, $\mathbf{E} $, and $\mathbf{H} $. $\mathbf{I} $ is inversion-odd ($P$-odd) and time-reversal odd ($T$-odd); $\mathbf{E} $ is $P$-odd and $T$-even; and $\mathbf{M} $ and $\mathbf{H} $ are $P$-even and $T$-odd. In accordance with these transformations, some response tensors must be zero when a system in a null field has $P$ and/or $T$ symmetry. For example, the linear electric-field-induced magnetization tensor $\alpha $ must be zero when a given system has a time-reversal symmetry because input $\mathbf{E} $ and output $\mathbf{M} $ acquire opposite signs through a time-reversal transformation.

\begin{figure}[t]
\centering
\includegraphics[]{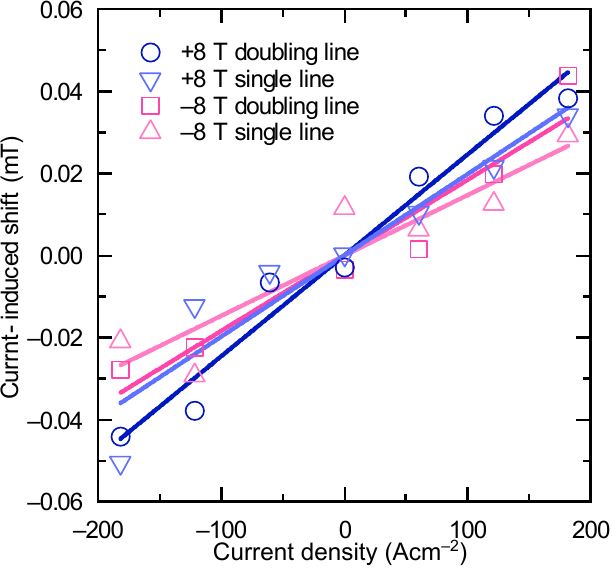}
\caption{
Current density dependence of the current-induced shifts for Sample \#1.
The current-induced shift is defined as the difference between the spectral first moment of each line and the shift origin defined such that the intersection of the fitting lines should be zero. The proportionality coefficients obtained by least-squares fitting are $+$2.5 ($\pm $0.2) $\times $ 10$^{-4} $ mTA$^{-1} $cm$^{2} $ for the doubling line ($+$8 T), $+$2.0 ($\pm $0.3) $\times $ 10$^{-4} $ mTA$^{-1} $cm$^{2} $ for the single line ($+$8 T), $+$2.0 ($\pm $0.3) $\times $ 10$^{-4} $ mTA$^{-1} $cm$^{2} $ for the doubling line ($-$8 T), and $+$1.8 ($\pm $0.3) $\times $ 10$^{-4} $ mTA$^{-1} $cm$^{2} $ for the single line ($-$8 T).
}
\label{Fig. 4} 
\end{figure}

For trigonal tellurium, $\alpha $ and $\Gamma ^{\mathrm{IH} } $ must be zero, and $\beta $ and $\Gamma ^{\mathrm{EH} } $ can be nonzero because tellurium is a nonmagnetic chiral (or, $T$-symmetric and $P$-broken) semiconductor. We can also neglect the effect of $\chi _{ij}H_{j}$, $\Gamma ^{\mathrm{IE}} _{i,jk} I_{j}E_{k}$, $\Gamma ^{\mathrm{II}} _{i,jk} I_{j}I_{k}$, $\Gamma ^{\mathrm{EE}} _{i,jk} E_{j}E_{k}$, and $\Gamma ^{\mathrm{HH}} _{i,jk} H_{j}H_{k}$ because the observed electrically induced NMR shift shows a linear dependence on an electric input, $\mathbf{I} $ or $\mathbf{E} $ (The linearity was reported in our previous work~\cite{Furukawa2017} and confirmed in Figs.~\ref{Fig. 4} and ~\ref{Fig. 6} of this work.) Thus, the possible origins of the electrically induced NMR shift are the linear current-induced magnetization term $\beta _{ij}I _{j}$ and/or the bilinear magnetoelectric term $\Gamma ^{\mathrm{EH}} _{i, jk}E_{j}H_{k}$. (Specifically, the bilinear magnetoelectric effect primarily originates from a modulation of the magnetic susceptibility induced by the piezoelectric effect.) We measured the magnetic-field polarity dependence of the electrically induced shift to clarify which effect dominates the electrically induced NMR shift observed in trigonal tellurium. The linear current-induced magnetization $\beta _{ij}I _{j}$ is independent of the polarity of a magnetic field; thus, the polarity reversal of a magnetic field does not change the polarity of the NMR shift induced by this effect. In contrast, the bilinear magnetoelectric $\Gamma ^{\mathrm{EH}} _{i, jk}E_{j}H_{k}$ term depends on the magnetic-field polarity; thus, the polarity reversal of the magnetic field reverses the polarity of the NMR shift induced by the bilinear magnetoelectric effect. 

Figures~\ref{Fig. 3}c--f show the current density dependence of the NMR spectra under the positive and negative magnetic fields of 8.012 T for Sample \#1, which has a left-handed [$P3_{2}21$($D_{3}^{6} $)] crystal structure (Fig.~\ref{Fig. 3}a). The polarity reversal of the magnetic field preserves the polarity of the observed electrically induced NMR shift. This experimental result demonstrates that the electrically induced NMR shift observed in trigonal tellurium is caused by the linear current-induced magnetization effect. Figure~\ref{Fig. 4} shows the current density dependence of the electrically induced NMR shift. The slope coefficients for the doubling line are $+$2.5 $\pm $ 0.2 $\times $ 10$^{-4} $ mTA$^{-1} $cm$^{2} $ for a positive field and $+$2.0 $\pm $ 0.3 $\times $ 10$^{-4} $ mTA$^{-1} $cm$^{2} $ for a negative field, showing a good agreement between the two values within the margin error. The slope coefficients for the single line are $+$2.0 $\pm $ 0.3 $\times $ 10$^{-4} $ mTA$^{-1} $cm$^{2} $ for a positive field and $+$1.8 $\pm $ 0.3 $\times $ 10$^{-4} $ mTA$^{-1} $cm$^{2} $  for a negative field, which also show a good agreement. The agreement between the coefficients for the positive and negative magnetic fields indicates that the bilinear magnetoelectric effect makes a negligible contribution to the electrically induced NMR shift compared to the linear current-induced magnetization effect.
Revisiting the symmetry consideration, the $\Gamma ^{\mathrm{EH}} _{i,zz}$ coefficients must be zero for the $32$($D_{3}$) point group symmetry of tellurium. In other words, all directional components of the bilinear EH-induced magnetization must be zero when external electric and magnetic fields are applied exactly parallel to the $c$-axis of tellurium. Although the present experimental setup had a slight misalignment of the magnetic and electric fields from the $c$ axis, this symmetry constraint guaranteed the weakness of the bilinear magnetoelectric effects for the present directions of the fields, even if the bilinear effect for the arbitrary field directions of an external fields would be detectable.

\subsection{Crystal structure chirality dependence of current-induced NMR shift}
\begin{figure}[t]
\centering
\includegraphics[]{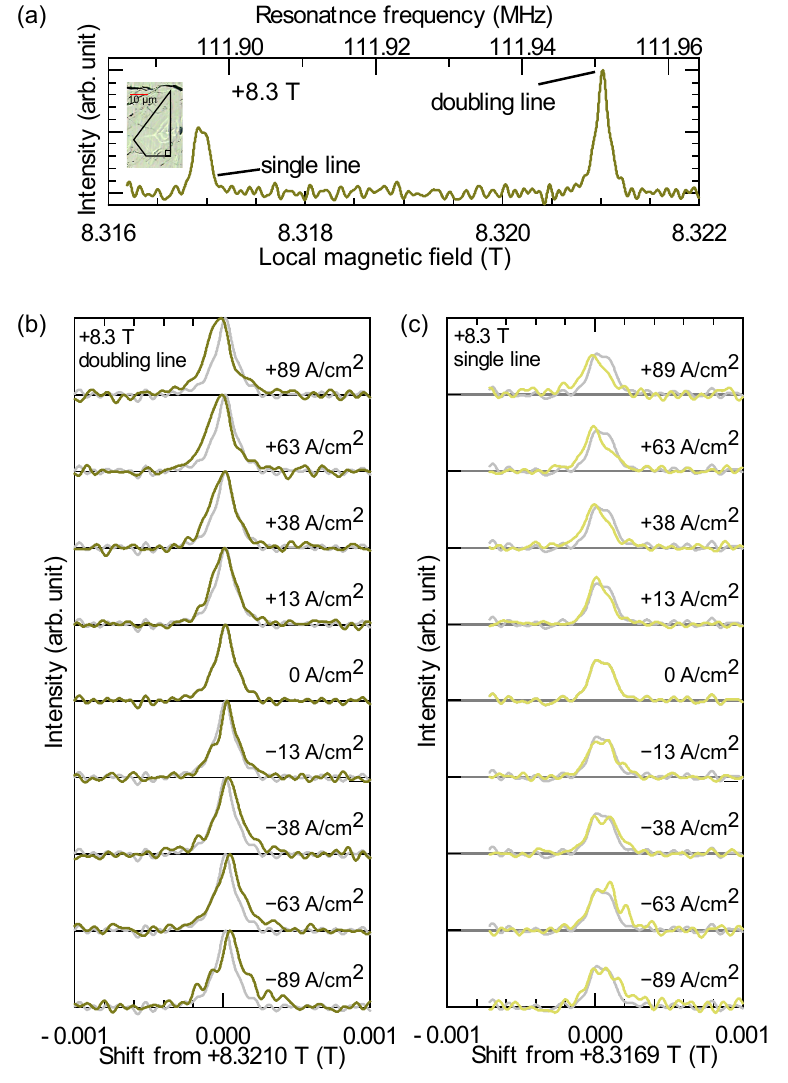}
\caption{
Inversion of the current-induced shift by a reversal of the crystal structure chirality .
(a) Single-crystal $^{125} $Te-NMR spectra for Sample \#2 in the absence of a pulsed electric current at 100 K. A magnetic field was applied approximately parallel to the $c$ axis, but slightly tilted to the $y$ axis. The spectra are plotted as a function of the effective magnetic field felt by the $^{125} $Te nuclei~\cite{SM}. The observed NMR spectra consist of a single line and a doubling line similar to the spectra for Sample \#1. The inset indicates the photo image of an etch pit on the crystal, whose type appears on the right-handed [$P3_{1}21$($D_{3}^{4} $)] crystal [27]. (b, c) Doubling (b) and single (c) lines in the $^{125} $Te-NMR spectrum for different electric current densities. For comparison, the gray lines in each row show the spectrum in the absence of a pulsed electric current.
}
\label{Fig. 5} 
\end{figure}
\begin{figure}[t]
\centering
\includegraphics[]{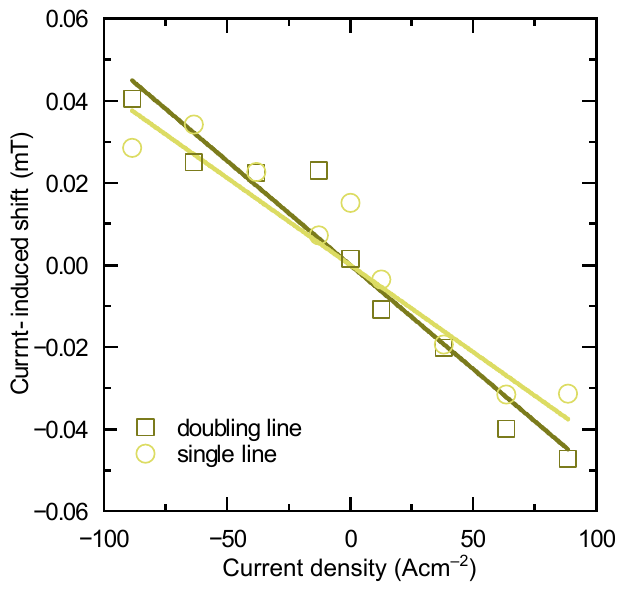}
\caption{
Current density dependence of the current-induced shifts for Sample \#2.
The current-induced shift is defined as the difference between the spectral first moment of each line and the shift origin defined such that the intersection of the fitting lines should be zero. The proportionality coefficients obtained by least-squares fitting are $-$5.1 ($\pm $0.4) $\times $ 10$^{-4} $ mTA$^{-1} $cm$^{2} $ for the doubling line and $-$4.2 ($\pm $0.5) $\times $ 10$^{-4} $ mTA$^{-1} $cm$^{2} $ for the single line..}
\label{Fig. 6} 
\end{figure}
We next show that the polarity of the current-induced magnetization in tellurium depends on the crystal structure chirality. Figures ~\ref{Fig. 5}b and c show the $^{125} $Te NMR spectra under a positive magnetic field of 8.327 T for Sample \#2, which has a right-handed [$P3_{1}21$($D_{3}^{4} $)] crystal structure (Fig.~\ref{Fig. 5}a). Similar to the results for Sample \#1 (Figs.~\ref{Fig. 3}c--f), an electric-current induces local-field shifts, and the shifts linearly depend on the electrical current density with coefficients of $-$5.1 $\pm $ 0.4 $\times $ 10$^{-4} $ mTA$^{-1} $cm$^{2} $ for the doubling line and $-$4.2 $\pm $ 0.5 $\times $ 10$^{-4} $ mTA$^{-1} $cm$^{2} $ for the single line (Fig.~\ref{Fig. 6}). Notably, the polarity of the shifts induced by a positive electric current was negative for Sample \#2 (right-handed tellurium) in contrast to the positive shifts for Sample \#1 (left-handed tellurium). The negative shifts indicate a negative spin magnetization because of the positive hyperfine coupling coefficient~\cite{Selbach1979} between the $c$-axis electronic spin magnetization and the $c$-axis hyperfine field. Therefore, the polarity of the current-induced spin magnetization is positive for the left-handed tellurium and negative for the right-handed tellurium when a positive current is applied. (We will discuss the possible current-induced orbital magnetization later.) Consequently, the present result shows that a chirality reversal of the crystal structure causes a reversal of the current-induced magnetization. The absolute values of the current-induced magnetization coefficients for the right- and left-handed samples differed by a factor of approximately two. This discrepancy may originate from a non-negligible amount of a surface contribution to the total electric current depending on the surface condition for each sample. If a surface current exists, a bulk current could be overestimated, and the coefficients could be underestimated. In fact, a metallic surface state in the $\{10\bar{1}0\}$ plane was observed~\cite{VonKlitzing1971,Averkiev1998,Akiba2020}. Note that even if a non-negligible amount of a surface current was present, the Oersted field generated by the surface current would cause only a broadening (not shift) of the NMR spectra; thus, the observed current-induced shift originates from the magnetization of the bulk part of the samples.

\section{Discussions}
\subsection{Relation among the current-induced NMR shift, k space spin texture, and crystal structure chirality}
We discuss the relation among the polarity of the current-induced NMR shift, spin texture of the highest valence bands, and crystal structure chirality. As discussed in our previous report~\cite{Furukawa2017}, an electric current induces magnetization parallel (antiparallel) to an electric current in tellurium with an inward (outward) spin-angular-momentum texture, which provides an outward (inward) spin-magnetization texture around the $H$ and $H'$ points. Thus, the present chirality-dependent result shows that a left-handed (right-handed) crystal has an inward (outward) spin-angular-momentum texture, which is consistent with the recent ab initio calculations and the spin- and angle-resolved photoemission spectroscopy measurements ~\cite{Tsirkin2018,Sakano2020}. We should point out herein that our previous report~\cite{Furukawa2017} had two mistakes. First, we wrote in the report that the crystal used was right-handed, but it was probably left-handed; and second, we wrote that a right-handed crystal has an inward spin-angular momentum texture, but it actually had an outward texture. The first mistake was caused by the possible multi-domain structure of the tellurium ingot used, which was grown by the Bridgman method without seed crystal. The second mistake was caused by the contradictions in the previous studies about the relations between the crystal structure chirality and the physical properties, such as natural optical activity, resonant diffraction with circularly polarized x rays~\cite{Tanaka2010,Tanaka2012}, and polarized neutron scattering~\cite{Brown1996}. We discuss these points in detail in Supplemental Material~\cite{SM}.

\subsection{Current-induced orbital magnetization}
The highest valence bands of tellurium have an orbital angular momentum that originates not from an intra-atomic orbital angular momentum of a $p$-wave atomic state, but from an inter-atomic orbital angular momentum of the Bloch state. Thus, an electric current can induce not only net spin magnetization, but also net orbital magnetization, which is called the orbital Edelstein effect ~\cite{Yoda2015,Yoda2018}. In fact, theoretical studies predicted current-induced intrinsic~\cite{Tsirkin2018,Sahin2018} and extrinsic~\cite{Sahin2018} orbital magnetization effect in trigonal tellurium. Unfortunately, we infer here that the observed NMR shift is not related to the current-induced orbital magnetization. Similar to spin magnetization, orbital magnetization also causes a hyperfine field at nuclei through a hyperfine coupling between the electronic orbital motion and the nuclear spins. We tried herein to estimate the orbital hyperfine coupling using the conventional chemical shift theory, in which the local electric current uniquely determines the hyperfine field at the nuclei through the Biot--Savart law. (Note that there might be room for argument about whether the conventional theory needs to be modified to be consistent with the modern theory of orbital magnetization~\cite{Thonhauser2005,Xiao2005}, in which the intrinsic orbital magnetization in a periodic system is represented not uniquely by the distribution of a microscopic current density, but by the geometrical properties of Hilbert space.) In this case, the orbital hyperfine coupling is inversely proportional to the cubic distance between the nucleus and the microscopic electric current of the Bloch state. The orbital magnetization of the Bloch state near the $H$ and $H'$ points of tellurium originates not from an intra-atomic current, but from an inter-atomic current like the solenoidal current along the tellurium chains. Thus, the effective distance between the microscopic current and the tellurium nuclei is long (of the order of the radius of the tellurium helix), yielding a very weak orbital hyperfine coupling. Moreover, the polarity of the observed current-induced shift also indicates that the current-induced orbital magnetization does not dominate the observed shift. The orbital hyperfine coupling constant must be positive according to the conventional electromagnetics. In addition, theoretical studies~\cite{Tsirkin2018,Sahin2018} showed that the intrinsic orbital magnetization of each Bloch state of the highest valence bands is antiparallel to the spin magnetization of each state. This suggests that a current-induced orbital magnetization should cause a negative NMR shift for the left-handed crystal under a positive current, whereas an observed shift is positive. The insensitiveness of the NMR to the current-induced orbital magnetization allows us to separate the spin and orbital contributions to the current-induced magnetization using the present NMR method and other probes that can detect the current-induced total magnetization.

\begin{acknowledgments}
We thank Koji Tanaka for experimental assistance.
This work was supported by JSPS KAKENHI Grant Numbers 17K14345, 18K03540, 19H02583, 19H01852 and Grant for Basic Science Research Projects from The Sumitomo Foundation.
\end{acknowledgments}

\bibliography{Tellurium2020}
\end{document}